\documentstyle[preprint,aps,eqsecnum]{revtex}
\begin{document}
\preprint{UCD 93-27, September 1993}
\title{
     Critical exponent for the density of percolating flux
   }
\author{Joe Kiskis}
\address{
Department of Physics\\
University of California, Davis, CA 95616, USA\\
jekucd@ucdhep.ucdavis.edu  }
\date{\today}
\maketitle
\begin{abstract}
This paper is a study of some of the critical properties of a simple model for
flux. The model is motivated by gauge theory and is equivalent to the Ising
model in three dimensions. The phase with condensed flux is studied. This is
the ordered phase of the Ising model and the high temperature, deconfined
phase of the gauge theory.  The flux picture will be used in this phase. Near
the transition, the density is low enough so that flux variables remain useful.
There is a finite density of finite flux clusters on both sides of the phase
transition. In the deconfined phase, there is also an infinite, percolating
network of flux with a density that vanishes as  $T \rightarrow T_{c}^{+}$.  On
both sides of the critical point, the nonanalyticity in the total flux density
is characterized by the exponent $(1-\alpha)$.  The main result of this paper
is a calculation of the critical exponent for the percolating network.  The
exponent for the density of the percolating cluster is
$ \zeta = (1-\alpha) - (\varphi-1)$. The specific heat exponent $\alpha$ and
the crossover exponent
$\varphi$ can be computed in the $\epsilon$-expansion. Since
$\zeta < (1-\alpha)$, the variation in the separate densities is much more
rapid than
that of the total. Flux is moving from the infinite cluster to the finite
clusters much more rapidly than the total density is decreasing.
\end{abstract}
\narrowtext

\section{Introduction}

This paper is a study of some of the critical properties of a simple model for
flux. The model is motivated by gauge theory and is equivalent to the Ising
model in three dimensions. The main emphasis is on the phase in which the flux
is condensed. This is the ordered phase of the Ising model and the high
temperature, deconfined phase of the gauge theory. In such a phase, there is a
percolating network of flux in addition to clusters of finite size.

The character expansion for the Ising model is a product over links.
Each term in the product has the form $(1+zss')$. The expansion of the product
can be viewed as a sum over configurations of links that have the $zss'$ term
rather than the 1. By definition, these are the links with flux.
This machinery is commonly used to calculate expansions
in the small $z$ phase. I will use it to describe configurations in the
large $z$ phase also. Naturally, it cannot be the basis for a perturbative
expansion in that phase. In the large $z$ phase, there is a percolating network
of links with flux. This network must disappear as  $z \rightarrow
z_{c}^{+}$. The main result of this paper is a calculation of the critical
exponent for the vanishing density of the percolating network.

All of the calculations are for the model introduced above. My interest in the
model originates in the subject of finite-temperature gauge theory.
In the low-temperature, confined phase
of a non-Abelian gauge theory, it is believed that sources are connected by a
tube of color-electric flux rather than by a dipole Coulomb field. This
flux-tube description of the low-temperature phase
has been used \cite{c1} to account for the properties of the second-order,
finite-temperature, deconfining phase transition of pure-glue, $SU(2)$, gauge
theory. At high
temperature, a plasma of free gluons with perturbative corrections can account
for some properties. However, there are also nonperturbative effects in the
deconfined phase. The correct physical picture for $T \rightarrow T_{c}^{+}$ is
not established. I will use the flux picture near the phase transition in the
deconfined phase. In this phase, the flux is condensed, but the density may be
low enough so that flux variables remain useful. I expect a finite density of
flux clusters of a finite size on both sides of the phase transition. In the
deconfined phase, there is also a percolating network of flux with a density
that vanishes as $T \rightarrow T_{c}^{+}$.

In the strong-coupling description of $SU(2)$ lattice gauge theory, the
flux on links is labeled by the irreducible representations of $SU(2)$. At
each site, the representations must combine to have a singlet piece. For
$SU(2)$, this can happen only if there is an even number of links with flux at
each site.

These configurations of flux on links are subject to both quantum and
finite-temperature fluctuations. The quantum fluctuations reflect the fact that
these configurations are eigenstates of the Hamiltonian only in the lowest
order of the strong coupling expansion. The magnetic term in the Hamiltonian
moves the flux on the lattice. Thus, the flux is thickened into tubes.  In the
usual, but unproven, picture of confinement, that is as far as it goes.
The flux is not spread into Coulomb-like configurations, and tubes of flux
remain a useful description. On a scale larger than the basic length
$1/\Lambda_{QCD}$, the eigenstates of the gauge field Hamiltonian are
configurations of flux tubes, and the energy is proportional to the length of
the flux. There is an additional assumption  that on these large scales, it is
sufficient to include only the basic flux strength associated with  fundamental
sources.

Finite-temperature fluctuations are the main interest of this work. As the
temperature increases, small loops and other clusters of flux appear with low
density. Increasing temperature brings increasing density. Most work \cite{c1}
with this kind of picture has concentrated on the region below the critical
temperature. At the critical temperature, there is a condensation of flux. I
assume that the flux picture can be carried into the high-temperature phase
where the condensation appears as a percolating network of flux. This paper
deals with the region just into the high-temperature phase.

With this background, the simplified model that will be studied
can be presented. Space is a three-dimensional cubic lattice.
There are variables $\theta$ on links that can have the values 1 or 0 to
indicate the presence or absence of flux. The energy of a link with flux is
$\sigma$. The weight for a configuration is
\begin{equation}
 e^{- \case{1}{T} \sum_{l} \sigma \theta(l)} =
            \prod_{l} e^{- \case{1}{T} \sigma \theta(l)}   .
\end{equation}
The sum over configurations is restricted to those in which the number of links
at a site that have flux is even. Let the collection of all such
configurations be $C'$. It is a subset of the unrestricted collection
of all configurations $C$. The partition function is
\begin{equation}
 Z = \sum_{C'} e^{- \case{1}{T} \sum_{l} \sigma \theta(l)} .
\end{equation}

Some ideas from percolation theory will be used to study this model. They are
introduced in the next section.
Also, it will be shown that the flux model is equivalent to the
Ising model. In the third section, the main result is the exponent for the
vanishing density of the percolating flux as $T \rightarrow T_{c}^{+}$.
\begin{equation}
 \rho_{\infty} \sim (T-T_c)^{\zeta}
\end{equation}
with
\begin{equation}
 \zeta = (1-\alpha) -(\varphi-1) .
\end{equation}
The specific heat exponent $\alpha$ and the crossover exponent $\varphi$ can be
computed in the \mbox{$\epsilon$-expansion}. The fourth section contains a
brief summary.

\section{Percolating Flux}

\subsection{General remarks}

The discussion of the previous section has led to a model
for flux defined by the partition function
\begin{equation}
 Z = \sum_{C'} e^{- \case{1}{T} \sum_{l} \sigma \theta(l)} .
\end{equation}
The link variable $\theta(l) = 1$ or 0 to indicate the presence or absence of
flux on link $l$.
The sum is restricted to configurations in which the number of links
at a site that have flux is even. In the high-temperature phase, there is a
condensation of flux. If the density of the condensed flux is not too high,
then a description in terms of flux variables will be useful.

To discuss this phase, I will draw on ideas from percolation
theory \cite{c5}.
Without the restriction on configurations, the model becomes pure bond
percolation with \mbox{$p = e^{-\sigma/T}/(1+e^{-\sigma/T})$} as
the probability for flux on a link.
In percolation, both phases have a finite density
of flux clusters of finite size. In the large $p$ phase,
there is also a single percolating cluster.
Since $p$ is the probability that a link has flux, $p$ is also the total
density of flux per link. Thus, in terms of the control parameter $p$, the
total density has exponent 1, i.e. $\rho_{total} = p_c + (p-p_c)^1$.
The critical value of $p_c$ is about $0.25$,
so the critical link density is also about $0.25$.
Since just a quarter of the links have flux, this remains a useful description
of configurations.

The density of the infinite cluster vanishes with
a rate given by the exponent $\beta \approx 0.403$.
It is important to note that the exponent $\beta$ is smaller than one. This
means that there must be compensating behavior in the density of finite
clusters for $p>p_c$. My interest is in the analogous relations for the
simplified flux model.

The flux model differs from percolation only through the restriction on
configurations. To study the flux model, the ideas from pure percolation can
be applied. In particular, assume that there is a single percolating cluster in
the high temperature phase and that the nonanalyticities at the phase
transition are characterized by critical exponents. Naturally, the  numerical
values of the exponents will be different from those of percolation.

\subsection{Equivalence with the Ising model}

Now it is time to show that the model for flux is equivalent to the Ising
model. To do this, introduce site variables to enforce the restriction
on configurations, and then do the $\theta$ sums. The site variables will
turn out to be the Ising spins.

First, consider the sum of $\theta(l)$
over the $l$'s contained in the set of links $L(i)$ that hit a site $i$:
\begin{equation}
  \sum_{l \in L(i)} \theta(l)  .
\end{equation}
To force this to be even at each site, introduce the site variables $s(i)$,
which take the values $\pm 1$. Then the factor
\begin{equation}
 \case{1}{2} \sum_{s(i)=\pm 1} s(i)^{\sum_{l \in L(i)} \theta(l)}
\end{equation}
has the desired effect. A factor like this is introduced into the
partition function sum for each site $i$. Then the restriction on the $\theta$
sums can be relaxed. With the abbreviations
\begin{equation}
 \sum_s \equiv \prod_i (\case{1}{2} \sum_{s(i)}) ,
\end{equation}
$z=e^{- \sigma/T}$, and $ss'$ for the product of the pair of spins
on the sites that bound a given link, the partition function is
\begin{eqnarray}
 Z & = & \sum_s \sum_{C} \prod_{l} e^{- \case{1}{T}  \sigma \theta(l)}
        \prod_i s(i)^{\sum_{l \in L(i)} \theta(l)} \\
   & = & \sum_s \sum_{C} \prod_{l}
            (z ss')^{\theta(l)}           \\
   & = & \sum_s \prod_l ( 1 + z ss')   .
\end{eqnarray}
This is a version of the Ising model
\begin{equation}
   \sum_s e^{\beta \sum_l ss'}
\end{equation}
with $z=\tanh \beta$.
The links with flux are those that receive the term
$zss'$ rather than 1 in the expansion of $\prod_l ( 1 + z ss')$.

\subsection{Additional remarks on $SU(2)$ gauge theory and the Ising model}

Thus, the simple model for $SU(2)$ flux is equivalent to the Ising model. The
appearance of the Ising model is not surprising because
the gauge theory and the Ising model are related by another route
\cite{c6}. Finite-temperature, $SU(2)$, gauge theory has a global $Z(2)$
symmetry. The order parameter for the second-order, gauge theory phase
transition is the Wilson line. The effective theory for the lines inherits the
$Z(2)$ symmetry. This effective theory is in the universality class of the
three-dimensional Ising model. An approximate, strong coupling, character
expansion for the effective theory involves a similar, but somewhat more
complicated, appearance of flux on links.

In Ref.\ \cite{c8}, there are discussions of the adjoint
Wilson line that involve the behavior of flux. The
finite-temperature expectation value of the adjoint line has the form \cite{r1}
\begin{equation}
 \langle L_{1} \rangle = c + d_{\pm} |\case{T-T_c}{T_c}|^{1-\alpha}   .
                                            \label{e11}
\end{equation}
In the deconfined phase, this receives contributions from the finite clusters
and from the percolating
cluster of flux. However, the separate contributions need not have the exponent
$(1-\alpha)$. This paper will address that
issue. To the extent that the simple model captures the essential aspects of
the gauge theory, then the conclusion will be that the two gauge exponents are
also different.

Now that the general picture has been outlined, the task of the next section is
to determine the behavior of the flux around the phase transition. The main
emphasis will be on the critical exponent for the density of the percolating
cluster.

\section{Calculations}

The purpose of this section is to calculate the critical exponents for the
total flux density and for the density of the percolating cluster. The
model has been formulated in terms of flux on links
\begin{equation}
 Z = \sum_{C'} e^{- \case{1}{T} \sum_{l} \sigma \theta(l)}
\end{equation}
and as the Ising model
\begin{equation}
 Z  =  \sum_s \prod_l ( 1 + z ss') .
\end{equation}
The discussion begins with the second form. Operators associated with the
desired flux densities must be identified. In order to separate the percolating
cluster, the spins will be generalized to unit vectors $\hat{s}$ that can point
along either direction of $N$ axes. Then the $N \rightarrow 1$ limit will be
taken to return to the original case. This is a technical device that is
closely related to a similar trick in \cite{c9}.

\subsection{Total flux density}

The total flux density can be obtained straightforwardly without
the generalization of variables mentioned above. Consider a
particular link, the spin variable pair $ss'$ on the sites that bound it,
and the operator $(z-ss')$. In the product over all links, this will
multiply the factor $(1+zss')$ for the link in question to give $(z^2-1)ss'$.
This eliminates configurations without flux on this link. Configurations with
flux on this link receive a factor of $(z^2-1)$ rather than $z$ in their
weight. Therefore, the expectation value of the operator
\begin{equation}
 X = \frac{z}{(z^2-1)}(z-ss')
\end{equation}
gives the probability that a given link has flux.

Evidently, $X$ is closely related to the internal energy per link
$E=-ss'$. Near the critical point
\begin{equation}
 -E = c + d_{\pm} |z-z_c|^{1-\alpha}      \label{e12}
\end{equation}
with $c \approx 0.332$, $z_c \approx 0.218$, and $\alpha \approx 0.12$
\cite{c10}.
Thus, the operator $X$ that gives the total flux density has the critical
value $X_c = z_c(z_c-c)/(z_c^2-1) = 0.0261$ and the critical exponent
$(1-\alpha) \approx 0.88$.

These numbers can be compared with analogs from pure percolation. At
the critical point, the factor for an additional link of flux is
$z_c \approx 0.22$. This is slightly larger than the analogous number $0.21$
for a self-avoiding walk \cite{cc1} and is somewhat smaller than the factor
$p_c/(1-p_c) \approx 0.33$ in pure percolation.
The critical density of flux in the Ising model is about one tenth
of what it is in pure percolation.
Since many clusters that are possible in percolation are not possible
in the Ising model, the total Ising density should be lower.
For example the smallest cluster in percolation has just one link, while in the
Ising model, it has four. With the same factor for an occupied link,
this will give a much
more dilute concentration of occupied links in the Ising model.
In the growth of a large cluster, the excluded volume will be smaller in the
Ising model so that the entropy per link of a large cluster will be larger in
the Ising model.
Since criticality occurs when the entropy and
energy balance, the critical value of
the link factor will be smaller in the Ising model than in percolation. This
further reduces the critical density in the Ising model relative to
percolation.

This concludes the initial discussion of the total flux density.

\subsection{Percolating flux density}

In the high temperature phase, the percolating cluster and the finite clusters
contribute to the total density. The goal is to determine the exponent for the
percolating cluster near the critical point where its density goes to zero.
It will turn out to be a bit smaller than the $(1-\alpha)$ for the
total density. It follows that there must be a
complementary increase in the contribution from the finite clusters. Thus, both
components vary more rapidly than their sum near the critical point. More flux
is moving from the percolating cluster to the finite clusters than is
disappearing.

The density of the percolating cluster can be separated by using a
somewhat indirect method.
Generalize the spins to unit vectors $\hat{s}$ that are restricted to point
in one of the $2N$ directions that are parallel or antiparallel to
the axes of an $N$-dimensional space. Generalize the partition function
to
\begin{equation}
 Z  =  \sum_s \prod_l ( 1 + z \hat{s}\cdot\hat{s}') .  \label{e1}
\end{equation}
Now $\sum_s$ means sum each spin over the $2N$ allowed orientations. This
defines the generalized spin model.

As before, the links with flux are those that have the term with the $zss'$
rather than 1 in the expansion of the product. All the spins in a cluster must
be along the same axis. Thus, there are $N$ mutually repelling flavors of Ising
flux.

In the $z>z_c$, condensed phase, an infinitesimal external field will determine
a preferred flavor and break the cubic symmetry.
I assume that there is a single percolating cluster on which the spins are
restricted to the preferred flavor. To be specific, let the preferred flavor be
the first spin component.

To measure the density of the percolating cluster, consider the spin $\hat{s}$
at a given site. It satisfies the simple identity
\begin{equation}
 \hat{s}^2 = \sum_a s_{a}^{2} = 1 .
\end{equation}
{}From the first spin component $s_1$, construct the operator
\begin{equation}
 Y =  s_{1}^{2} - \case{1}{N}   .
\end{equation}
Its average will be zero in the symmetric phase. In the percolating phase,
the finite clusters are still symmetric, so they do not contribute to
$\langle Y \rangle$.
On sites hit by the percolating cluster, $Y = (N-1)/N$. Thus,
\begin{equation}
 \langle Y \rangle = \case{N-1}{N} \times \text{(site density of the
percolating
                                                        cluster)}
\end{equation}
There are a number of equivalent forms for $Y$:
\begin{eqnarray}
 Y & = & s_{1}^{2} - \case{1}{N} \\
   & = & s_{1}^{2} - \case{1}{N}\sum_a s_{a}^{2} \\
   & = & \case{N-1}{N} s_{1}^{2} - \case{1}{N}\sum_{a \neq 1} s_{a}^{2} \\
   & = & \case{N-1}{N} ( s_{1}^{2} - s_{2}^{2} )            .
\end{eqnarray}
The last equality is true if we limit ourselves to matrix elements of $Y$
with other operators that contain $s_1$ only. This is sufficient for our
purposes. The expectation value of the operator
$\bar{Y} \equiv (s_{1}^{2} - s_{2}^{2})$ gives the density of the percolating
cluster.

Recall that the goal is to find the exponent $\zeta$ in
\begin{equation}
 \langle \bar{Y} \rangle \sim (T-T_c)^{\zeta} .
\end{equation}
For $(T-T_c)<0$ where there is no percolating cluster,
$\langle \bar{Y} \rangle =0$.
The exponent is related to the anomalous dimension of the operator in the usual
manner \cite{c11}
\begin{eqnarray}
 \zeta & = & \nu d_{\bar{Y}} \\
       & = &  \nu ( d-2 + \gamma_{\bar{Y}} ) \\
       & = &  (1-\alpha) - (\varphi -1)  .
\end{eqnarray}
The combination $\varphi = \nu ( 2 - \gamma_{\bar{Y}})$ is a crossover
exponent \cite{cc2}.
It will turn out that $\varphi > 1$. Thus, the decrease in
the density of the percolating cluster is much more rapid
than that of the total density.

Note that the operator $\bar{Y}$ gives the density of {\em sites} hit by the
percolating cluster. Earlier, it was link densities that appeared.
The ratio of sites hit to links hit is certainly between $1/3$ and 1.
Also, the density of the percolating cluster approaches zero. It follows that
the critical exponents for the percolating site and link densities must be the
same. For the finite clusters, where the density does not approach zero, the
site and link exponents might be different.

The task at hand is to calculate the anomalous dimension of $\bar{Y}$
and then take
$N \rightarrow 1$. To do this, I will pass to a field theory description and
use the $\epsilon$-expansion.

\subsection{Field theory}

The next step is to find a field theory version of the generalized spin model
in which the anomalous dimension of $\bar{Y}$ can be computed. First, ideas
from the renormalization group and universality will be used as the basis for a
guess. Then a systematic transformation and approximation will be done to show
that the guess can be put on a stronger foundation.

A field theory in the universality class of the spin model should
be on a three-dimensional, Euclidean space and should have a field with the
internal symmetry of the order parameter. The generalized model has
$N$-dimensional cubic symmetry. A simple guess is
$N$-component, scalar field theory with ``cubic anisotropy'' \cite{c12}.
The Lagrangian density is
\begin{equation}
   {\cal L} = \case{1}{2} (\partial \phi)^2 + \case{1}{2} m^2 \phi^2 +
                 \case{g}{4!} (\phi^2)^2 +
                 \case{g'}{4!} \sum_a \phi_{a}^4    \label{e7}
\end{equation}
We also need an operator to play the role of $\bar{Y}$.
An obvious
combination to try is $\phi_{1}^{2} - \phi_{2}^{2}$. This can be related to the
more general and much-studied \cite{cc2}
\begin{equation}
 B = \frac{1}{N} [  (N-M)\sum_{a=1}^{M} \phi_{a}^{2} -
                   M \sum_{a=M+1}^{N} \phi_{a}^{2}  ]  .   \label{e8}
\end{equation}
We are interested in the specialization $M \rightarrow 1$
followed by $N \rightarrow 1$ at the end. If we
stick to matrix elements with $\phi_1$, then
\begin{equation}
 B \rightarrow \case{N-1}{N} (\phi_{1}^{2} - \phi_{2}^{2} ) .
\end{equation}
Besides the evident relation to $Y$ and $\bar{Y}$, the operator in (\ref{e8})
has satisfying
technical properties. It does not mix with $\phi^2$ under renormalization,
so that it has its own anomalous dimension. Also, its anomalous dimension is
independent of $M$.

To put this guess in a more respectable position, a generalized Gaussian
transformation \cite{c13} can be done. First, the partition function for the
generalized model must be put in the form of a sum over configurations with a
Boltzmann weight $e^{-H/T}$. Then a conventional path to a field theory can be
followed. Since the model does not satisfy
$(\hat{s} \cdot \hat{s}')^2 = 1$, the Hamiltonian must be more complicated than
$H \propto \sum_l \hat{s} \cdot \hat{s}'$. The two-term form
\begin{equation}
 H/T = \sum_l [ a  \hat{s} \cdot \hat{s}' + b (\hat{s} \cdot \hat{s}')^2  ]
                            \label{e2}
\end{equation}
is sufficient. For the partition function, this gives
\begin{equation}
 Z = \sum_s \prod_l e^{-[a  \hat{s} \cdot \hat{s}' +
                           b (\hat{s} \cdot \hat{s}')^2 ] } .  \label{e9}
\end{equation}
If this is to agree with (\ref{e1}), we need
\begin{equation}
 z=-\tanh a  \; \;  \text{ and } \; \;   e^b = \cosh a  .   \label{e10}
\end{equation}

For the Gaussian transformation, a slightly different expression for
summand in (\ref{e2}) is a better starting point \cite{r2}:
\begin{equation}
 a \sum_a s_a s'_{a} + b \sum_a (s_a)^2 (s'_a)^2     .  \label{e14}
\end{equation}
Now follow the method described in Ref.\ \cite{c13}. This involves the
introduction of two pairs of $N$-component, real-valued fields.
The pair $u_a$ and $\phi_a$ is associated with $s_a$, and the pair $v_a$
and $\chi_a$ is associated with $s_a^2$.
In the end, there will be a field theory for $\phi$ alone.

First $Z$ is rewritten as
\begin{equation}
 Z = \int\!\!d\phi\!\int\!\!d\chi\!\int\!\!du\!\int\!\!dv \sum_s
        e^{-\sum_l (a \vec{u}\cdot\vec{u}' + b \vec{v}\cdot\vec{v}')}
    \prod_i e^{\sum_a \{\phi_a(i)[u_a(i)-s_a(i)]+\chi_a(i)[v_a(i)-s_a^2(i)]\}}
,
\end{equation}
then the sum on the spin variables $s$ is done. This gives
\begin{eqnarray}
 Z = \int\!\!d\phi\!\int\!\!d\chi\!\int\!\!du\!\int\!\!dv\,
       && e^{-\sum_l (a \vec{u}\cdot\vec{u}' + b \vec{v}\cdot\vec{v}')
   +\sum_i [ \vec{\phi}(i)\cdot\vec{u}(i) + \vec{\chi}(i)\cdot\vec{v}(i) ] }
                                                           \nonumber \\
    && (\prod_i \sum_{a_{i}}) (\prod_j e^{-\chi_{a_{j}}(j)})
     [\prod_k 2\cosh \phi_{a_{k}}(k)]   .      \label{e6}
\end{eqnarray}
The next step is to do the $\chi$ and $v$ integrals. These are in a factor
that is a function of the indices $\{ a_{i} \}$
\begin{eqnarray}
 Z_1 & = & \int\!\! d\chi\! \int\!\! dv\,
        e^{-\sum_l b \vec{v}\cdot\vec{v}'
           +\sum_i [\vec{\chi}(i)\cdot\vec{v}(i) - \chi_{a_{i}}(i) ] }  \\
     & = & e^{-b \sum_l \delta(a_i,a_{i'})} .
\end{eqnarray}
The $\delta(\cdot,\cdot)$ is the Kronecker delta. The subscripts $i$ and $i'$
refer to the sites that bound the link $l$. Notice that this is the
Boltzmann weight for an $N$-state Potts model with site variables $a_i$.
This could also have been obtained directly from (\ref{e14}).
Next include the $a_i$ sums and the $\cosh$ factors to get a functional of
$\phi$.
\begin{eqnarray}
 Z_2 & = & (\prod_i \sum_{a_{i}}) e^{-b \sum_l \delta(a_i,a_{i'})}
                     [\prod_k 2\cosh \phi_{a_{k}}(k)]  \\
     & = & Z_P \langle\!\langle \prod_k 2\cosh \phi_{a_{k}}(k)
                                                      \rangle\!\rangle
\end{eqnarray}
The quantities $Z_P$ and $\langle\!\langle \cdot\rangle\!\rangle$ refer to the
partition function
and expectation value for the Potts model.

Note that this is an antiferromagnetic Potts model. It is in the
disordered phase. Equation (\ref{e10}) shows that, when $z$ and $a$ are small,
$b \approx a^2/2$. So $b$
will be much smaller than $a$, and the second term in (\ref{e2}) is a small
correction that can be computed in a strong-coupling series.
Without the second term in (\ref{e2}), the mean field estimate is
$a_c \approx N/6$. This gives $b \approx 0.13$ for $N=3$, and $b \approx 0.056$
for $N=2$. For $N=1$, the actual $a_c \approx 0.22$ can be used to give
$b \approx  0.024$. These are small numbers. For comparison, consider that the
critical values for the antiferromagnetic Potts model are $b_c \approx 0.82$
for $N=3$
and $b_c \approx 0.44$ for $N=2$ \cite{c13a}. The $N=1$ case is trickier. The
$N=1$, ferromagnetic Potts model is percolation. The critical probability
$1-e^{b_c} = p_c \approx 0.25$ corresponds to $b_c \approx -0.29$. The $N=1$,
antiferromagnetic case is percolation with negative probability.
The power series in $p$ that gives $p_c \approx 0.25$ has all positive
coefficients through order $p^{10}$. However, an analysis of the series
indicates the presence of a weak singularity for negative $p$ that is a little
closer to the origin \cite{c13b}. There is no evidence for a singularity
that would correspond to $b$ as small as $0.024$.

The contribution that $Z_2$ makes to the
action for $\phi$ will be computed through order $\phi^4$.
In the end, it will be easy to see that higher powers of $\phi$ are irrelevant.

After expanding to order $\phi^4$, computing the Potts expectation values, and
exponentiating, the $\phi$-dependent part of the exponent is
\begin{equation}
 -U = \sum_i \{ \langle\!\langle [ \case{1}{2}\phi_{a_i}^2(i)
                 + \case{1}{24}\phi_{a_i}^4(i) ] \rangle\!\rangle
               -\case{1}{2} \langle\!\langle \case{1}{2}\phi_{a_i}^2(i)
                                                        \rangle\!\rangle^2 \}
\\
 +\case{1}{2}\sum\sum_{i \neq j} \langle\!\langle \case{1}{2}\phi_{a_i}^2(i)
                          \case{1}{2}\phi_{a_j}^2(j) \rangle\!\rangle_c  .
\end{equation}
The subscript $c$ on the last Potts average indicates that it is connected.
Using the fact that all values of $a_i$ are equally likely in the symmetric
phase, the first three terms become
\begin{equation}
  \sum_i \{ \case{1}{N}\sum_a [\case{1}{2}\phi_a^2(i)
                 + \case{1}{24}\phi_a^4(i) ]
      -\case{1}{2} [ \case{1}{N}\sum_a\case{1}{2}\phi_a^2(i)]^2 \} .
                                              \label{e5}
\end{equation}
The summand of the last term in $-U$ becomes
\begin{equation}
 \case{1}{4} \sum_{ab} \phi_a^2(i) \phi_b^2(j) D_{ab}(i,j) \label{e3}
\end{equation}
where
\begin{equation}
 D_{ab}(i,j) \equiv \langle\!\langle \delta(a,a_i) \delta(b,b_j)
                                                       \rangle\!\rangle_c
\end{equation}
is the connected Potts two-point function.

With the Potts model away from its critical point, $D$ is short-ranged.
Since it is the
long-wavelength part of $\phi$ that determines its critical behavior,
we can take $\phi$ to be constant over the range of $|i-j|$ for
which $D$ is significantly different from zero. The first correction
is an irrelevant operator with
spatial derivatives and four powers of $\phi$.
That leaves
\begin{equation}
 \sum_i \case{1}{4} \sum_{ab} \phi_{a}^2(i) \phi_{b}^2(i)
             \case{1}{2} \sum_{j \neq i} D_{ab}(i,j)      \label{e4}
\end{equation}
for the last term in $-U$.
The internal index structure of $D_{ab}$ has the symmetry of
the symmetric phase of the Potts model. This allows for a term that is
independent of the indices $a$ and $b$ and for a term that is proportional to
$\delta(a,b)$. Thus, the terms that appear in (\ref{e4}) are of the same
form as those in (\ref{e5}). The only effect of the term
$b (\hat{s} \cdot \hat{s}')^2$ is to alter the coefficients of the $\phi^4$
terms in the potential.

Now we see that the contribution from $U$ to the action for $\phi$
is a $\phi^4$ potential with cubic anisotropy. Returning to (\ref{e6}), we have
\begin{equation}
 Z = \int\!\! d\phi\!\int\!\!du\,
        e^{-\sum_l a \vec{u}\cdot\vec{u}'
           +\sum_i  \vec{\phi}(i)\cdot\vec{u}(i)  -U }  .
\end{equation}
The last step is to do the $u$ integral. However, there is little point in
carrying the calculation further. It is clear that, except for the
$b$-dependence of the coefficients in $U$, this effective theory for $\phi$ is
the same as that which would have come from the conventional, bilinear form
\mbox{$a(\hat{s}\cdot\hat{s}')$}. It is a textbook result \cite{c13} that
the $u$ integral and another long-wavelength approximation give a ``kinetic''
term for $\phi$ and a $\phi^2$ term to add to $U$.

The conclusion of this discussion is that a Gaussian transformation confirms
the original guess. The field theory associated with the generalized spin model
is an $N$-component, $\phi^4$, scalar field theory with cubic anisotropy. Since
this is a well-studied example \cite{c12}, we can draw from established
results.

\subsection{Fixed points and exponents}

Now that the preliminaries are out of the way, we can get to the essentials
relatively easily. The heavy lifting has already been done by others and is
described rather nicely in \cite{c12}. For this field theory, there are five
possibilities revealed by the $\epsilon$-expansion. There can be a
fluctuation-induced, first-order phase transition. Since the Ising
and the $SU(2)$ phase transitions are both second order, this case
can be eliminated.
The second possibility is the Gaussian fixed point. This has free field
exponents and is unstable in both directions in the coupling
constant plane. Both these properties make it unlikely to be the case of
interest. The bare action of (\ref{e7}) begins away from the origin of the
coupling plane. The Ising and gauge models have nontrivial fixed points.

There are three nontrivial fixed points called Ising, cubic, and Heisenberg. At
the Ising fixed point, the $N$ components decouple into $N$ independent fields
each at the 1-component $\phi^4$ fixed point. There is one stable and one
unstable direction. Again, it is unlikely that (\ref{e7}) will hit it.
It has already been noted that the Ising flavors of the generalized spin model
repel each other.
Equivalently, (\ref{e7})
includes coupling that is unstable at this fixed point. However,
(\ref{e7}) is an approximation, and there could be some fine cancellation that
has been spoiled. If this were the correct
fixed point, then the anomalous dimensions for $\phi^2$ and $B$ would be the
same, and the exponents for the percolating and the finite clusters would be
the same.

The cubic fixed point has the original cubic symmetry, while the Heisenberg
fixed point has full $O(N)$ symmetry. The relative stability of these two
depends upon the value of $N$. For our case with $N \rightarrow 1$, the
Heisenberg point is stable, while the cubic one has one stable and one unstable
direction. Again, in the absence of some special cancellations that the
approximations have missed, the cubic fixed point is not the one of interest.
Also, in the $N \rightarrow 1$ limit, this fixed point joins the Gaussian fixed
point. The anomalous dimension for $\phi^2$ is zero. Since this is not the
correct behavior for the total density, this cannot be the correct fixed
point.

Finally, we come to the stable and, therefore, most likely case: the
Heisenberg fixed point. The $\epsilon$-expansions for the anomalous dimensions
of $\phi^2$ and $B$ are in Ref.\ \cite{cc2}. The lowest order
results for $N \rightarrow 1$ lead to
\begin{equation}
 \alpha = \epsilon / 6 \text{\hspace*{.3cm}and\hspace*{.3cm}}
                                            \varphi - 1 = \epsilon / 18 .
\end{equation}
Since $ \zeta =   (1-\alpha) - (\varphi -1)$, this gives the exponent for the
percolating flux. For $\epsilon=1$, the result is
$ \zeta = (1-\alpha) - 1/18$. This may not be particularly precise. However, it
reveals the important qualitative fact that $\zeta$ is less than
$(1-\alpha)$.

\section{Summary}

In the low temperature phase, there are only finite clusters. In the high
temperature phase, the total flux is the sum of the finite clusters and the
percolating cluster. On both sides of the critical point, the nonanalyticity in
the total flux density is characterized by the exponent $(1-\alpha)$:
\begin{equation}
 \rho_{total} = c + d_{\pm} |T-T_c|^{1-\alpha}  .   \label{e13}
\end{equation}
The main result of this work is the exponent for the
vanishing density of the percolating flux as $T \rightarrow T_{c}^{+}$.
\begin{equation}
 \rho_{\infty} \sim (T-T_c)^{\zeta}
\end{equation}
with
\begin{equation}
 \zeta = (1-\alpha) -(\varphi-1) .
\end{equation}
The specific heat exponent $\alpha$ and the crossover exponent $\varphi$ can be
computed in the \mbox{$\epsilon$-expansion}.

Since $\zeta < (1-\alpha)$, the variation in the separate densities is much
more
rapid than that of the total. Flux is moving from the infinite cluster to the
finite clusters much more rapidly than the total density is decreasing.

\acknowledgments

I am grateful to R. Singh for helpful remarks.
It is a pleasure to acknowledge the hospitality of the Institute of
Theoretical Dynamics at UC Davis where some of this work was done.
This research was supported by the United States Department of Energy.


\begin{references}
\bibitem{r1} In (\ref{e11}), (\ref{e12}), and (\ref{e13}), the $c$ and
$d_{\pm}$ are generic constants that are not necessarily equal across
equations. The $\pm$ subscript on $d_{\pm}$ indicates that the amplitude can
have different values on the two sides of the phase transition.
\bibitem{r2} I hope it is clear that the Hamiltonian parameters $a$ and $b$ are
           distinct from the spin component indices $a$ and $b$.
\bibitem{c13a} The $N=3$ value is from
                 J.-S. Wang, R. Swendsen, and R. Kotecky,
                    Phys. Rev. B {\bf 42}, 2465 (1990).
               The $N=2$ value is the same as $\beta_c \approx 0.22$ for the
                   antiferromagnetic Ising model.
\end{references}
\end{document}